\documentclass[aps,twocolumn,pra,superscriptaddress,showpacs,showkeys,floatfix]{revtex4-1}

\usepackage{amsbsy}
\usepackage{graphicx}
\usepackage{amsmath}
\usepackage{amssymb}
\usepackage{bm}
\usepackage{tabularx}
\usepackage{booktabs}
\input{epsf}

\begin{document}

\title{Reconstruction of electromagnetic field states by a probe qubit}

\author{Fabrizio Angaroni}
\affiliation{Center for Nonlinear and Complex Systems,
Dipartimento di Scienza e Alta Tecnologia,
Universit\`a degli Studi dell'Insubria, via Valleggio 11, 22100 Como, Italy}
\affiliation{Istituto Nazionale di Fisica Nucleare, Sezione di Milano,
via Celoria 16, 20133 Milano, Italy}
\author{Giuliano Benenti}
\affiliation{Center for Nonlinear and Complex Systems,
Dipartimento di Scienza e Alta Tecnologia,
Universit\`a degli Studi dell'Insubria, via Valleggio 11, 22100 Como, Italy}
\affiliation{Istituto Nazionale di Fisica Nucleare, Sezione di Milano,
via Celoria 16, 20133 Milano, Italy}
\author{Giuliano Strini}
\affiliation{Department of Physics, University of Milan,
via Celoria 16, 20133 Milano, Italy}

\begin{abstract}
We propose a method to measure the quantum state of a single mode of
the electromagnetic field. The method is based
on the interaction of the field with a probe qubit. The  
qubit polarizations along coordinate axes 
are functions of the interaction time 
and from their Fourier transform we can in general fully reconstruct
pure states of the field and obtain partial information
in the case of mixed states. The method is illustrated 
by several examples, including the superposition of Fock states,
coherent states, and exotic states generated by the dynamical
Casimir effect.  
\end{abstract}



\maketitle

\section{Introduction}
\label{sec:intro}

The state $\rho$ of a quantum system encodes all information
that can be obtained on the system, namely the probabilities of all 
measurement outcomes are inferred from the quantum state. 
Being a statistical concept, an a priori unknown quantum state cannot
be obtained from a single measurement, but can instead be reconstructed 
through measurements on an ensemble of identically prepared copies
of the same state $\rho$. Such state reconstruction is known as quantum state
estimation or quantum tomography. The problem of determining a state $\rho$
of the system from measurements on multiple copies goes back to 
Fano~\cite{Fano}, who called \emph{quorum} a set of observable sufficient
to fully reconstruct the state. For a $d$-dimensional system, the 
density matrix is determined by $d^2-1$ independent parameters~\cite{footnote} 
and therefore 
$d-1$ projective measurements are necessary to determine 
such parameters, since measuring one observable can give only 
$d-1$ parameters~\cite{Fano}. For $d=2$ we can reconstruct the state of
a qubit from the polarization measurements along three coordinate axes 
$x$, $y$, and $z$. 
For an electromagnetic field mode, the quorum consists of a collection 
of quadrature operators measured through balanced homodyne detection, 
$X_\theta= (\cos\theta) Q + (\sin\theta) P$, with 
$Q$ and $P$ position and momentum of the harmonic oscillator representing
the field mode. Since a quantum harmonic oscillator has infinite dimension,
strictly speaking and infinite quorum is needed (i.e., infinite values of 
the continuous parameter $\theta$), but a finite quorum is, under certain assumptions, 
in practice sufficient to reconstruct the density 
matrix~\cite{Dariano1,Paris,Dariano2,Lvovsky,Manko}. 
In spite of the fact that quantum state tomography is an old 
problem~\cite{Fano}, interest in the field is still growing, 
mainly due to the development of quantum technologies for 
precision measurements, quantum communications, quantum cryptography,
and quantum computing, all applications for which a reliable state
determination is crucial. 

\begin{figure}
\begin{center}
\centerline{\includegraphics[scale=0.22]{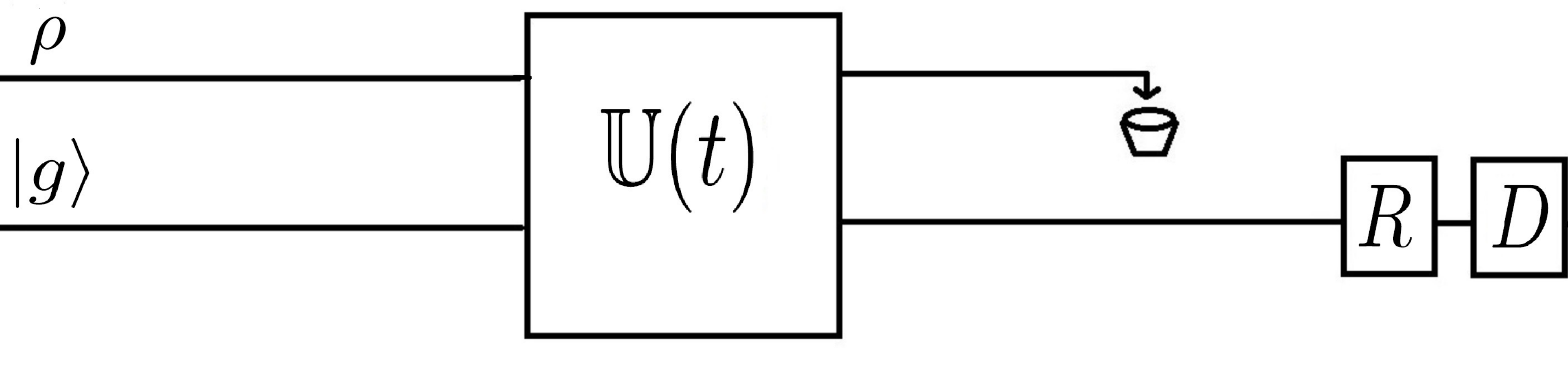}}
\caption{Schematic drawing of the stroboscopic tomographic method 
discussed in the text. After the field-qubit interaction up to 
time $t$, a rotation of the qubit state (represented by a box with a 
R inside) allows to detect the qubit polarization
along a selected direction (measurement represented by a box
with a D inside).}
\label{fig:stroboscopic}
\end{center}
\end{figure}

In this paper, we propose a new method for the partial reconstruction
of the state of a single mode of the electromagnetic field.
As usual in quantum tomography, we suppose that we can prepare 
repeatedly the system in the same state and we wish to obtain 
information about such \emph{target state} by means of the 
measurement of the expectation values of a suitable 
set of observables.
Our approach 
(see Fig.~\ref{fig:stroboscopic} for a schematic drawing)
is based on the \emph{coherent} interaction of the 
field with a \emph{probe qubit},
supposed to be weak enough to be treated 
within the rotating wave approximation. Assuming that we know
the dynamics of the overall field plus qubit state, then we 
can use a \emph{stroboscopic} approach. That is, from the 
measurements at different times of the qubit polarizations
along three coordinate axes and from the Fourier transform
of the mean values of such measurements we can obtain 
partial information on the target field state. 
As we shall discuss in detail below, we can reconstruct the
diagonal and the superdiagonal (or equivalently the 
subdiagonal) of the state. Such information is in general 
sufficient for a full reconstruction in the case of 
pure states, since from the diagonal elements we can reconstruct
the populations of the state and from the superdiagonal the 
relative phases. 

Our paper is organized as follows. In Sec.~\ref{sec:method}
we describe our tomographic method, which is illustrated 
in Sec.~\ref{sec:examples} by several examples: Fock states,
superposition of Fock states, coherent states and states 
generated by the dynamical Casimir effect. 
Statistical errors due to finite number of measurements
are discussed in Sec.~\ref{sec:errors}.
Our conclusions are drawn in Sec.~\ref{sec:conc}.

\section{Method}
\label{sec:method}

Our state reconstruction method is based on the 
interaction of a single-mode of the quantized electromagnetic field
with a probe qubit. The overall field-qubit system is described 
by the Jaynes-Cummings hamiltonian~\cite{micromaser}
\begin{equation}
\begin{array}{c}
{\displaystyle
H=H_0+H_I,
}
\\
\\
{\displaystyle
H_0=
\omega\left(a^\dagger a +\frac{1}{2}\right)
-\frac{1}{2}\,\omega_a \sigma_z,
}
\\
\\
{\displaystyle
H_I=g \,\sigma_+a
+g^\star \sigma_-a^\dagger,
}
\end{array}
\label{eq:JaynesCummings}
\end{equation}
where we set the reduced Planck's constant $\hbar=1$, 
$\sigma_i$ ($i=x,y,z$) are the Pauli matrices,
$\sigma_\pm = \frac{1}{2}\,(\sigma_x\mp i \sigma_y)$
are the rising and lowering operators for the qubit:
$\sigma_+ |g\rangle = |e\rangle$,
$\sigma_+ |e\rangle = 0$,
$\sigma_- |g\rangle = 0$,
$\sigma_- |e\rangle = |g\rangle$;
the operators $a^\dagger$ and $a$ for the field create
and annihilate a photon:
$a^\dagger |n\rangle=\sqrt{n+1}|n+1\rangle$,
$a |n\rangle=\sqrt{n}|n-1\rangle$,
$|n\rangle$ being the Fock state with $n$ photons.
For the sake of simplicity, we consider a real 
coupling strength, $g\in\mathbb{R}$, and the resonant case,
$\omega=\omega_a$.

The Jaynes-Cummings model is exactly solvable: in the 
$\{|g,0\rangle,|e,0\rangle,|g,1\rangle,|e,1\rangle,
|g,2\rangle,|e,2\rangle,...\}$ basis we can write 
the time evolution operator $\mathbb{U}(t)$ up to time $t$ in a
block diagonal form:
\begin{equation}
 \mathbb{U}(t)=
 \begin{pmatrix}
  1&0&0&0 &\dots \\
  0&U^{(1)}(t)&0&0&\dots \\
  0&0&U^{(2)}(t)& 0&\dots\\
  0& 0&0&U^{(3)}(t)& \dots\\
  \vdots&\vdots&\vdots&\vdots&\vdots
 \end{pmatrix}.
\label{eq:UJC}
\end{equation}
Here the interaction picture has been used, 
the top left matrix element equal to one indicates 
that the $|g,0\rangle$ state evolves trivially, while 
the $2\times 2$ matrices 
$$
    U^{(n)}(t) =
    \begin{pmatrix}
 U_{11}^{(n)}(t) &  U_{12}^{(n)}(t)\\
 U_{21}^{(n)}(t) &  U_{22}^{(n)}(t)
    \end{pmatrix}
$$
   \begin{equation}
   = \begin{pmatrix}
      \cos(\Omega_{n}t) & -i\sin(\Omega_{n}t)\\
  -i\sin(\Omega_{n}t) & \cos(\Omega_{n}t)
    \end{pmatrix}
   \end{equation}
describe coherent Rabi oscillations between the atom-field
states $|g,n\rangle$ and $|e,n-1\rangle$, with 
the Rabi frequencies $\Omega_n=g\sqrt{n}$ ($n=1,2,3,...$).
 
Our purpose is to obtain information on a generic target state of 
the field $\rho=\sum_{i,j=0}^\infty
\rho_{ij} |i\rangle\langle j|$.
We assume that the qubit is prepared
in its ground state: $\rho^{(q)}=|g\rangle\langle g|$, so that 
the overall field-qubit state
reads $\rho_{\rm tot}=\rho^{(q)}\otimes\rho=
 |g\rangle\langle g|\otimes \rho$. 
By evolving such state up to time $t$ under the Jaynes-Cummings 
Hamiltonian, we have 
$\rho_{\rm tot}(t)=\mathbb{U}(t)\rho_{\rm tot}\mathbb{U}^\dagger$.
Tracing over the field we obtain $\rho^{(q)}(t)={\rm Tr}_f[\rho_{\rm tot}(t)]$.
By means of Eq.~(\ref{eq:UJC}) we can easily write the matrix elements
of $\rho^{(q)}(t)$ as follows:
$$
\rho^{(q)}_{gg}(t)=
\rho_{00}+\sum_{n=1}^\infty U_{22}^{(n)}(t)\rho_{nn}
[U_{22}^{(n)}(t)]^\star
$$
\begin{equation}
=\rho_{00}+\sum_{n=1}^\infty \rho_{nn} \cos^2(\Omega_n t),
\label{eq:rhogg}
\end{equation}
$$
\rho^{(q)}_{ge}(t)=
\rho_{01}[U^{(1)}_{12}(t)]^\star
+\sum_{n=1}^\infty U_{22}^{(n)}(t)\rho_{n, n+1}
[U_{12}^{(n+1)}(t)]^\star
$$
\begin{equation}
=i [\rho_{01}\sin(\Omega_1 t)
+\sum_{n=1}^\infty \rho_{n, n+1} \cos(\Omega_n t)\sin(\Omega_{n+1} t)].
\label{eq:rhoge}
\end{equation}

From the polarization measurements of the probe qubit along the 
coordinate axes at time $t$ 
we obtain the Bloch sphere coordinates $x(t),y(t),z(t)$, simply 
related to the matrix elements of 
$\rho^{(q)}(t)$~\cite{qcbook}: 
\begin{equation}
\rho^{(q)}_{gg}(t)=\frac{1+z(t)}{2},
\quad
\rho^{(q)}_{ge}(t)=\frac{x(t)-iy(t)}{2}.
\label{eq:Bloch}
\end{equation}
\begin{widetext}
From Eqs.~(\ref{eq:rhogg}), (\ref{eq:rhoge}), 
and (\ref{eq:Bloch}) we obtain
\begin{equation}
z(t)=\rho_{00}+\sum_{n=1}^{\infty}\rho_{nn}\cos(2\Omega_nt),
\end{equation}
\begin{equation}
x(t)=-\{2\,{\rm Im}(\rho_{01})\sin(\Omega_1 t) +
\sum_{n=1}^\infty {\rm Im}(\rho_{n,n+1})
[\sin((\Omega_{n+1}+\Omega_n)t)+\sin((\Omega_{n+1}-\Omega_n)t)]\},
\end{equation}
\begin{equation}
y(t)=-\{2\,{\rm Re}(\rho_{01})\sin(\Omega_1 t) +
\sum_{n=1}^\infty {\rm Re}(\rho_{n,n+1})
[\sin((\Omega_{n+1}+\Omega_n)t)+\sin((\Omega_{n+1}-\Omega_n)t)i]\}.
\end{equation}
Finally, the Fourier transforms~\cite{noteFourier} of $x(t)$, $y(t)$, and $z(t)$ 
are given by
\begin{equation}
\tilde{z}(\omega)=\rho_{00}\delta(\omega)+
\frac{1}{2}\,\sum_{n=1}^\infty
\rho_{nn}[\delta(\omega-2\Omega_n)+\delta(\omega+2\Omega_n)],
\label{eq:zw}
\end{equation}
$$
\tilde{x}(\omega)=i\{{\rm Im}(\rho_{01})[\delta(\omega+\Omega_1)-
\delta(\omega-\Omega_1)]
$$
\begin{equation}
+\frac{1}{2}
\sum_{n=1}^\infty {\rm Im}(\rho_{n,n+1})
[\delta(\omega+(\Omega_{n+1}+\Omega_n))-\delta(\omega-(\Omega_{n+1}+\Omega_n))
+\delta(\omega+(\Omega_{n+1}-\Omega_n))
-\delta(\omega-(\Omega_{n+1}-\Omega_n))]\},
\label{eq:xw}
\end{equation}
$$
\tilde{y}(\omega)=i\{{\rm Re}(\rho_{01})[\delta(\omega-\Omega_1)+
\delta(\omega+\Omega_1)]
$$
\begin{equation}
+\frac{1}{2}
\sum_{n=1}^\infty {\rm Re}(\rho_{n,n+1})
[\delta(\omega+(\Omega_{n+1}+\Omega_n))-\delta(\omega-(\Omega_{n+1}+\Omega_n))
+\delta(\omega+(\Omega_{n+1}-\Omega_n))
-\delta(\omega-(\Omega_{n+1}-\Omega_n))]\}.
\label{eq:yw}
\end{equation}
\end{widetext}

Hence from the Fourier transform of $z(t)$ we
can obtain the populations $\rho_{nn}$ of the target 
(initial) state of the field, while from the 
Fourier transform of $x(t)$ and $y(t)$ we can 
reconstruct the coherences $\rho_{n,n+1}$ ($n=0,1,...$). 
This information is in general sufficient to fully reconstruct
a pure state 
\begin{equation}
|\psi\rangle=\sum_{n=0}^\infty c_n |n\rangle=\sum_{n=0}^\infty
|c_n|e^{i\phi_n}|n\rangle.
\end{equation} 
Indeed in this case 
$|c_n|=\sqrt{\rho_{nn}}$, while
\begin{equation} 
\rho_{n,n+1}=c_nc_{n+1}^\star=|c_n||c_{n+1}|e^{i(\phi_n-\phi_{n+1})}. 
\label{eq:relativephases}
\end{equation}
Knowing $|c_n|$ and $|c_{n+1}|$ from the populations 
$\rho_{nn}$ and $\rho_{n+1,n+1}$ 
(i.e., from the measurements of the qubit polarization along $z$),
we can derive the relative phase 
$\phi_{n,n+1}\equiv \phi_n-\phi_{n+1}$, and therefore
fully reconstruct the state $|\psi\rangle$.
This method for pure states only fails in the particular cases 
where there exist two coefficients $c_j,c_l\ne 0$ and in between a
vanishing coefficient $c_k=0$ ($j<k<l$), as, for instance, in a 
state like $|c_0||0\rangle+|c_2|e^{i\phi_{02}}|2\rangle$. In this
case, our method only allows us to determine the populations
$|c_0|,|c_2|$ but not the relative phase $\phi_{02}$. 

\section{Examples}
\label{sec:examples}

We illustrate the general method of Sec.~\ref{sec:method}
in a few examples ordered by growing complexity.
Note that the equations reported below are instances of
the general expressions 
(\ref{eq:zw})-(\ref{eq:yw}).

\subsection{Fock states}

In this case $\rho=|n\rangle\langle n|$ and
$\tilde{z}(\omega)=\frac{1}{2}\,[\delta(\omega-2\Omega_n)+
\delta(\omega+2\Omega_n)]$ for $n\ne 0$, while 
$\tilde{z}(\omega)=\delta(\omega)$ for $n=0$. 
On the other hand, $\tilde{x}(\omega)=
\tilde{y}(\omega)\equiv 0$ as coherences are zero
for this state. 

\subsection{Superposition of Fock states}

Let us consider two particular cases, showing that the method
is useful to reproduce not only the populations but also the 
coherences of a state.
We consider $\frac{1}{\sqrt{2}}\,(|1\rangle+|2\rangle)$ (state $1$)
and $\frac{1}{\sqrt{2}}\,(|1\rangle+e^{i\pi/4}|2\rangle)$ (state $2$).
Let $(x_i,y_i,z_i)$ denote the Bloch vector for state $i$ ($i=1,2$).
The two states have the same populations and 
$$
\tilde{z}_1(\omega)=\tilde{z}_2(\omega)=
\frac{1}{4}\,
[\delta(\omega-2\Omega_1)+\delta(\omega+2\Omega_1)
$$
\begin{equation}
+\delta(\omega-2\Omega_2)+\delta(\omega+2\Omega_2)].
\end{equation}
The first state is real and therefore $\tilde{x}_1(\omega)=0$, while
$$
\tilde{y}_1(\omega)=
\frac{i}{4}\,
[\delta(\omega+(\Omega_1+\Omega_2))-\delta(\omega-(\Omega_1+\Omega_2))
$$
\begin{equation}
+\delta(\omega+(\Omega_2-\Omega_1))-\delta(\omega-(\Omega_2-\Omega_1))].
\end{equation}
On the other hand, for the second state ${\rm Re}(\rho_{12})=
{\rm Im}(\rho_{12})$ and we have 
$\tilde{x}_2(\omega)=\tilde{y}_2(\omega)=\frac{\sqrt{2}}{2}\,
\tilde{y}_1(\omega)$.

\begin{widetext}

\begin{figure}
\begin{center}
\centerline{\includegraphics[scale=0.45]{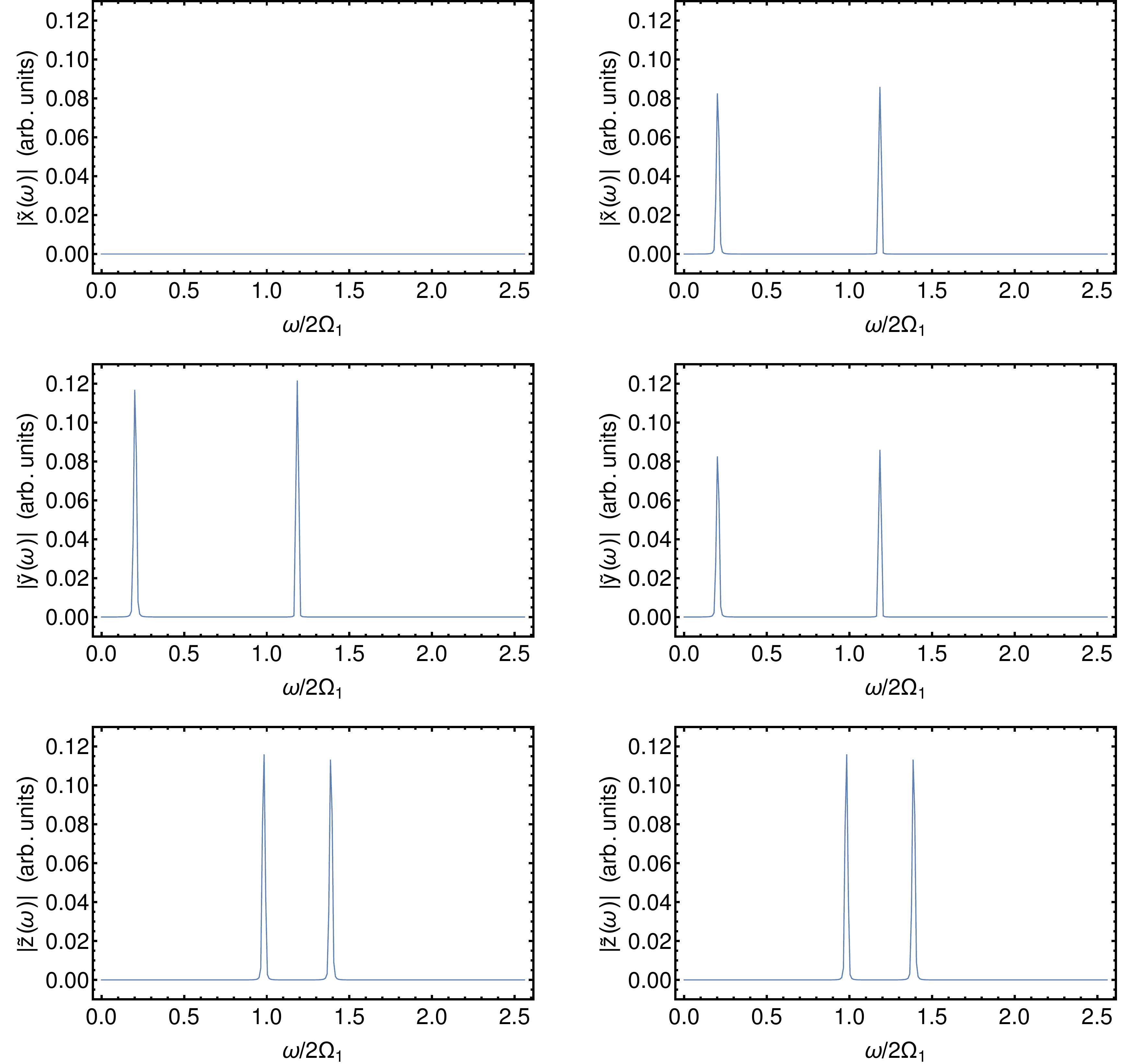}}
\caption{Fourier transforms $\tilde{x}(\omega)$,
$\tilde{y}(\omega)$, and $\tilde{z}(\omega)$ of the Bloch
sphere coordinates of the probe qubit, for the field input states
$\frac{1}{\sqrt{2}}\,(|1\rangle + |2\rangle)$ (left)
and $\frac{1}{\sqrt{2}}\,(|1\rangle + e^{i\pi/4}|2\rangle)$ (right).
Note that we only show the Fourier transforms for $\omega>0$,
since they are symmetric for $\omega\to-\omega$.}
\label{fig:superposition}
\end{center}
\end{figure}
\end{widetext}

These results are illustrated in Fig.~\ref{fig:superposition}, 
where we simulate 
an experiment repeated for $N_t=4096$ interaction times $t_k$, 
separated by a time
step $\Delta t=0.075/\Omega_1$, assuming that for each time $t_k$
an ideally unlimited number of experimental runs is performed
(the impact of statistical errors due to finite number of 
measurements will be discussed in Sec.~\ref{sec:errors}).
Due to the finite maximum interaction time $T= N_t (\Delta t)$,
the delta functions of the above expressions are broadened. 
Nevertheless, by measuring the overall area below the peaks
we can reconstruct with very good accuracy 
(three significant digits) the non-zero matrix elements of the 
density operator $\rho$ for the two states,
see the table below. For instance, by adding the areas below the peaks at 
$\omega=2\Omega_1$ and $\omega=-2\Omega_1$ in $\tilde{z}_1=\tilde{z}_2$ we
obtain $\rho_{11}=0.5004$, to be compared with 
the exact value $\rho_{11}=\frac{1}{2}$.  
For the state $\rho_2$, we obtain from $\tilde{x}_2$ and $\tilde{y}_2$
that ${\rm Re}[\rho_{12}]={\rm Im}[\rho_{12}]=0.3532$,
to be compared with the exact value ${\rm Re}[\rho_{12}]={\rm Im}[\rho_{12}]=
1/2\sqrt{2}\approx 0.3536$. 
\begin{center}
\begin{table}[h!]
\begin{tabular}{|c|c|c|}
\hline
\toprule
 & state $1$  & state $2$   \\
\hline
\midrule
$\rho_{11}$     & 0.5004            &  0.5004  \\
$\rho_{22}$ &        0.4997          &  0.4997 \\
Re$[\rho_{12}]$&    0.4998             &   0.3532  \\
  Im$[\rho_{12}]$ & 0                &  0.3532\\
\bottomrule
\hline
\end{tabular}
\label{tab1}
\end{table}
\end{center}

\subsection{Coherent states}

\begin{figure}
\begin{center}
\centerline{\includegraphics[scale=0.5]{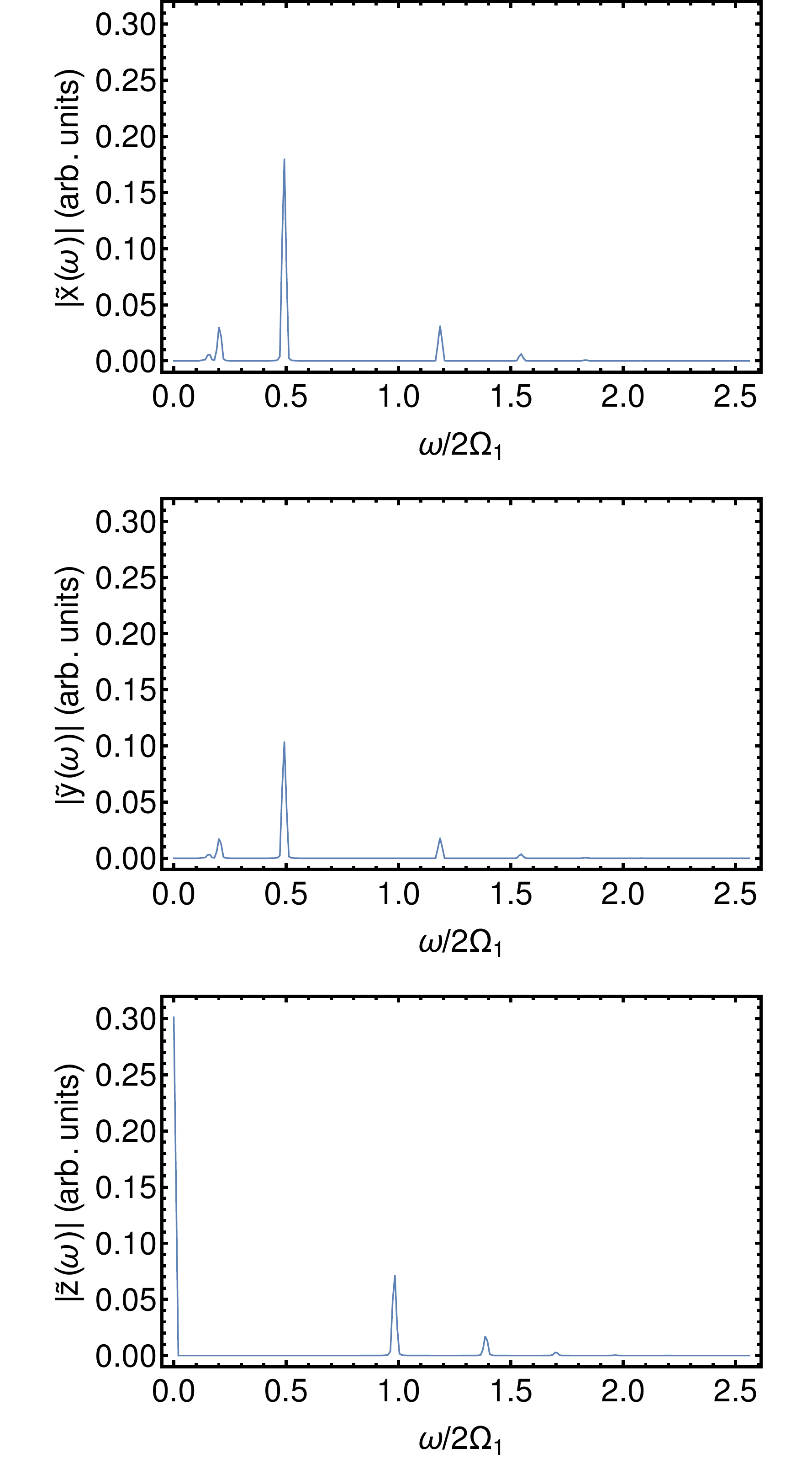}}
\caption{As in \ref{fig:superposition} but for a coherent state
with $\alpha=0.7e^{i\pi/3}$.}
\label{fig:coherent}
\end{center}
\end{figure}

\begin{figure}
\begin{center}
\centerline{\includegraphics[scale=0.75]{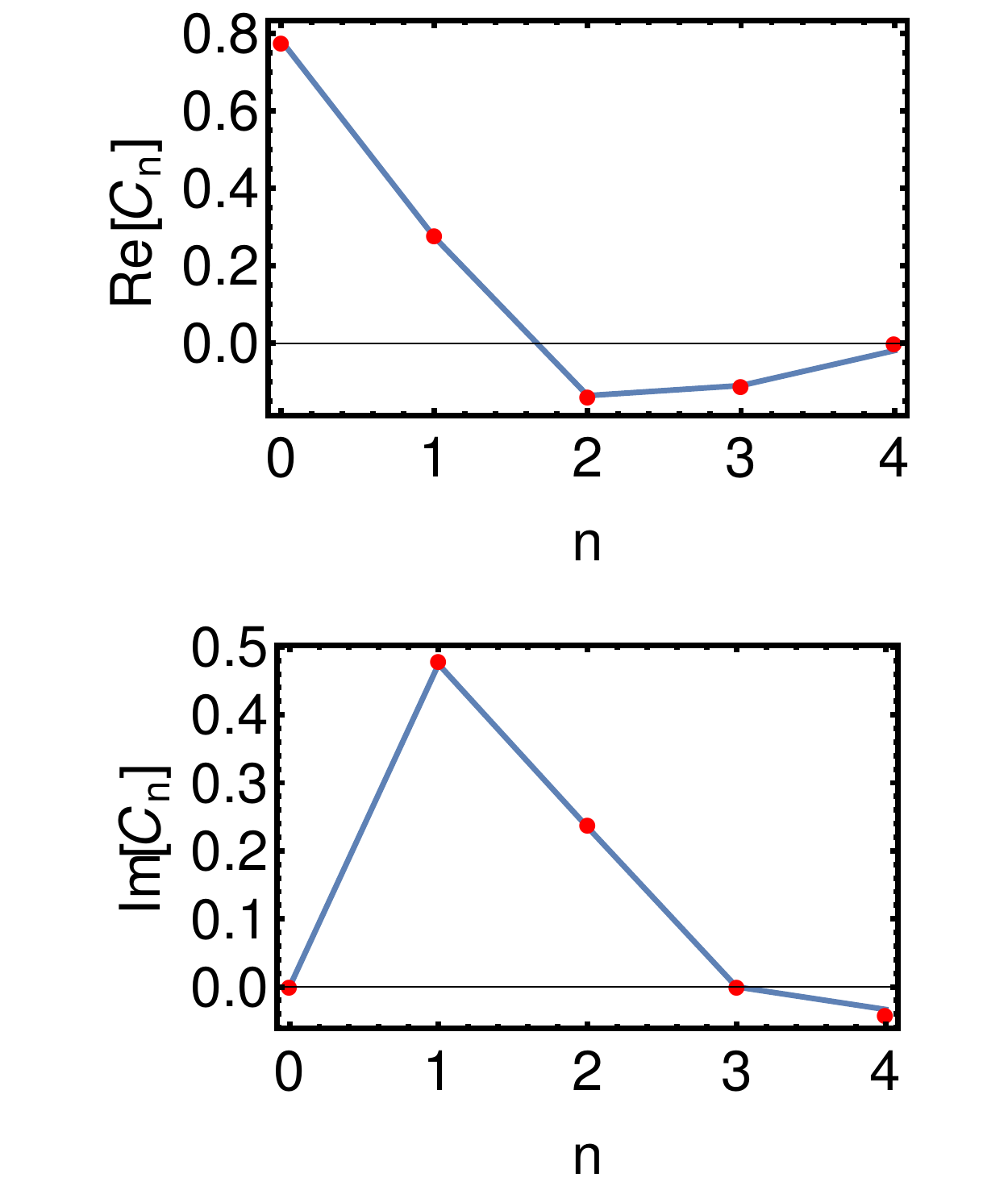}}
\caption{Real and imaginary part of a coherent state 
with $\alpha=0.7e^{i\pi/3}$ (line) and state reconstruction 
by means of the Fourier transforms $\tilde{x}(\omega)$,
$\tilde{y}(\omega)$, and $\tilde{z}(\omega)$ of Fig.~\ref{fig:coherent}
(circles).}
\label{fig:reimcoherent}
\end{center}
\end{figure}

We now consider a coherent state, whose representation in the basis of
Fock states reads
$|\alpha\rangle =\sum_{n=0}^\infty c_n |n\rangle$, where 
$c_n= \exp\left(-\frac{|\alpha|^2}{2}\right)\frac{\alpha^n}{\sqrt{n !}}$, with 
$\alpha\in\mathbb{C}$. 
In the simulations reported below we use a complex value 
$\alpha=0.7e^{i\pi/3}$ to demonstrate that not only populations
but also coherences of the field can be measured. 
In the bottom panel of Fig.~\ref{fig:coherent} 
we can clearly see the peaks at 
$\omega=2\Omega_n$ for $n=0,1,2,3$, corresponding to 
$\omega/2\Omega_1=0,1,\sqrt{2},\sqrt{3}$, respectively. By integrating the 
areas below these peaks (and the symmetric peaks at $\omega=-2\Omega_n$ 
not shown here) we reconstruct the populations $\rho_{nn}$.
The coherences $\rho_{n,n+1}$ are then obtained from 
the areas below the peaks in 
$\tilde{x}(\omega)$ and $\tilde{y}(\omega)$ in Fig.~\ref{fig:coherent}.
The relative phases $\phi_n-\phi_{n+1}$ are then derived via
Eq.~(\ref{eq:relativephases}). Finally, all 
the phases $\phi_n$ are obtained once the overall arbitrary phase
is set, for instance we can choose $\phi_0=0$.
Knowing $|c_n|=\sqrt{\rho_{nn}}$ and $\phi_n$, we have fully reconstructed
the field state. The good agreement between the results obtained by
means of our tomographic method and the exact field state is shown in 
Fig.~\ref{fig:reimcoherent}.

\subsection{Exotic states from the dynamical Casimir effect}

The Dynamical Casimir Effect (DCE)~\cite{moore,dodonov,noriRMP}
is the generation of photons from the vacuum due to time-dependent
boundary conditions for the electromagnetic field.
Such quantum vacuum amplification effect
has been observed in experiments with superconducting
circuits~\cite{norinature,lahteenmaki},
and also investigated in the context of
Bose-Einstein condensates~\cite{jaskula}, in exciton-polariton
condensates~\cite{koghee}, for multipartite entanglement generation
in cavity networks~\cite{solano2014,savasta}, for quantum
communication protocols~\cite{casimirqip}, 
for quantum technologies~\cite{adesso} and also in 
the context of finite-time quantum thermodynamics~\cite{frigo}.

The field state we want to reconstruct by means of our tomographic protocol 
is obtained from the 
interaction between a two-level system and a single mode of
the quantized electromagnetic field, described by the
time-dependent Rabi Hamiltonian~\cite{micromaser}
\begin{equation}
  \begin{array}{c}
{\displaystyle
H(t)=H_0+H_I(t),
}
\\
\\
{\displaystyle
H_0=
\omega\left(a^\dagger a +\frac{1}{2}\right)
-\frac{1}{2}\,\omega_a \sigma_z ,
}
\\
\\
{\displaystyle
H_I(t)=f(t)\,[\,g \,\sigma_+\,(a^\dagger+a)
+g^\star \sigma_-\,(a^\dagger+a)\,],
}
\end{array}
\label{eq:noREWAquantum}
\end{equation}
where we assume sudden switch on/off of the coupling:
$f(t)=1$ for $0\le t \le \tau$, $f(t)=0$ otherwise
\cite{footparametric}. 
It is the non-adiabatic switching of the interaction
that leads to the DCE, namely to the generation of photons
even though initially both the field and the qubit are in their
ground state. Hamiltonian (\ref{eq:noREWAquantum}) leads
to the generation of \emph{exotic states} of the field
with negative components in their Wigner function~\cite{exotic}.
Such states are very different from the squeezed
states obtained in approximate descriptions of the DCE via 
quadratic Hamiltonians~\cite{bastard,noriRMP}.

We consider as initial condition the state
$|\Psi_0\rangle=|g,0\rangle$ and 
compute numerically the qubit-field state after the
interaction time:
$|\Psi(\tau)\rangle = c_g(\tau)|g\rangle|\phi_g(\tau)\rangle+
c_e(\tau)|e\rangle|\phi_e(\tau)\rangle$, where
$|\phi_g\rangle$ and $|\phi_e\rangle$
are normalized states of the field.
Note that we define the initial state $|g,0\rangle$
before switching on the interaction and consider
the final state $|\Psi(\tau)\rangle$ after the interaction
has been switched off.
By changing the interaction strength $g$ and the
interaction time
$\tau$ we can generate a great variety of states of the
field~\cite{exotic}, both in the unconditional case
and in the conditional
case in which the final qubit state is measured, for
instance in the $\{|g\rangle,|e\rangle\}$ basis.
In the first case, the field state reads
$\rho={\rm Tr}_q |\Psi\rangle\langle\Psi|=
|c_g|^2|\phi_g\rangle\langle\phi_g|
+|c_e|^2|\phi_e\rangle\langle\phi_e|$, where
${\rm Tr}_q$ denotes the trace over the qubit subsystem;
in the latter case, we obtain the (pure) states
$\rho_g=|\phi_g\rangle\langle\phi_g|$ or
$\rho_e=|\phi_e\rangle\langle\phi_e|$.

\begin{figure}
\begin{center}
\centerline{\includegraphics[scale=0.078]{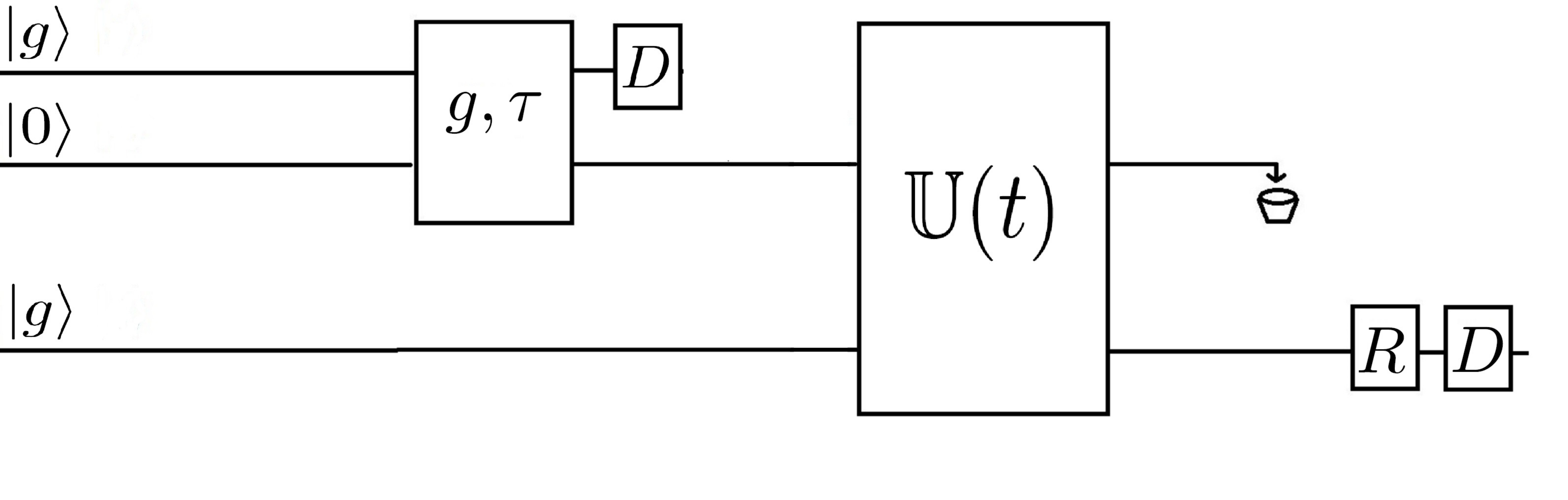}}
\caption{State reconstruction of field states generated by
the dynamical Casimir effect, with field-qubit interaction 
strength $g$ and interaction time $\tau$. Pure states of the field
are then obtained after measurement of the qubit polarization
and analyzed via the state reconstruction method described in 
this paper, with $\mathbb{U}(t)$ time evolution operator 
corresponding to the Jaynes-Cummings interaction of the field 
with a probe qubit up to time $t$. }
\label{fig:DCEtomography}
\end{center}
\end{figure}

\begin{figure}
\begin{center}
\centerline{\includegraphics[scale=0.6]{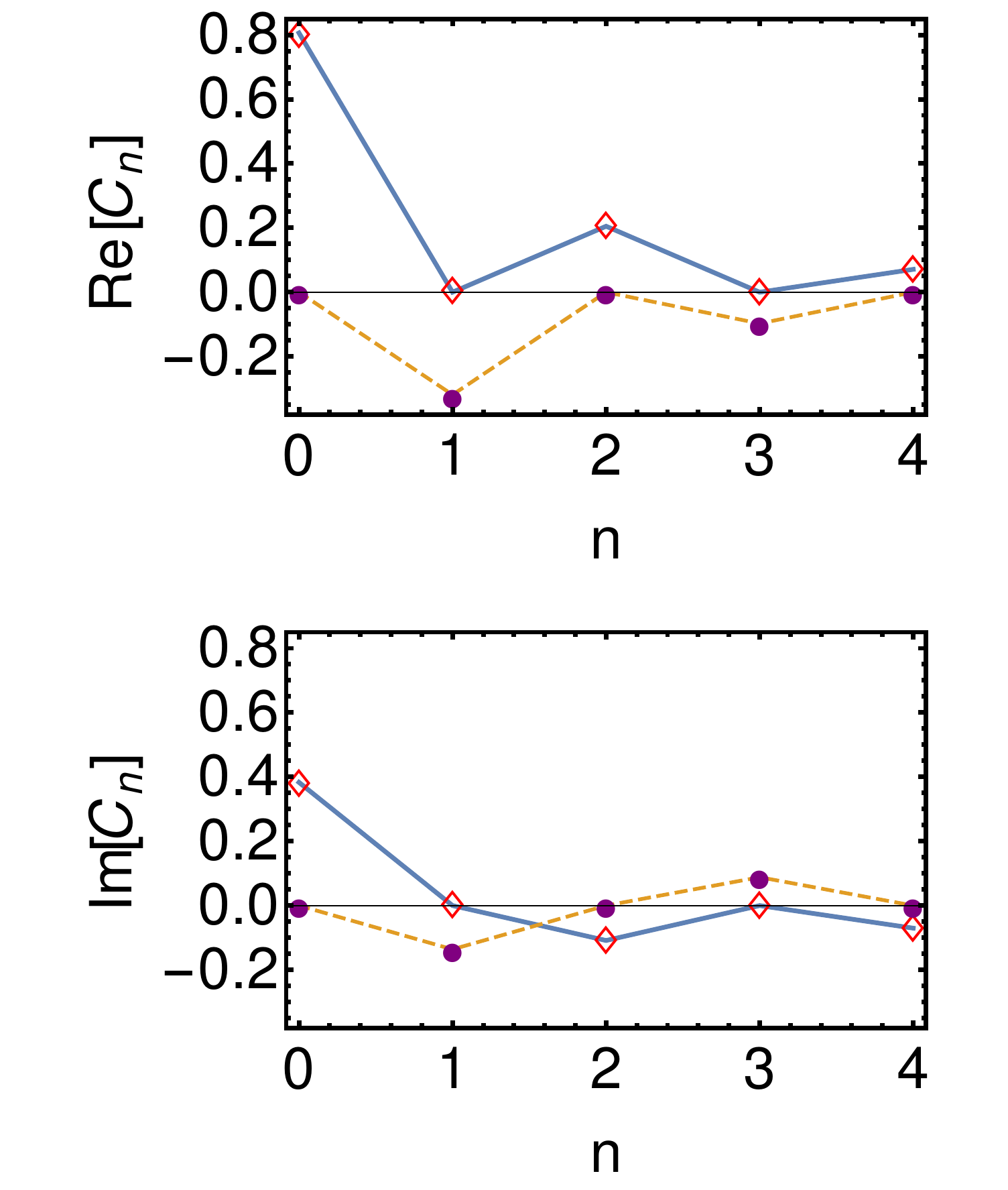}}
\caption{Real and imaginary part of the states 
$|\phi_g(\tau)\rangle$ (full line and diamonds) 
and $|\phi_e(\tau)\rangle$ 
(dashed line and circles) in the dynamical Casimir effect 
with $\omega=\omega_a$, 
$g/\omega=0.5$ and interaction time $\tau=\pi/2g$. 
Lines stand for the exact results,
symbols for the data obtained by state reconstruction.}
\label{fig:DCE0}
\end{center}
\end{figure}

As illustrated in Fig.~\ref{fig:DCEtomography}, 
the states $\rho_g$ and $\rho_e$ can be reconstructed 
via the tomographic method introduced in this paper. 
There is, however, a catch: since the parity 
$\Pi=\sigma_z (-1)^{a^\dagger a}$ of the excitations
is conserved by the Rabi Hamiltonian, 
for the conditional states $\rho_g$ and $\rho_e$
only the Fock states with respectively an even and an odd number of
photons are populated.
We are therefore in the situation in which
in the expansions of $|\phi_g\rangle$ and $|\phi_e\rangle$
on the Fock basis there exist vanishing 
coefficients between non-zero coefficients and
therefore our tomographic method would fail.
However, this problem can be overcome if the conditional 
states are obtained after measuring in the
$\{|+\rangle,|-\rangle\}$ basis, with 
$|\pm\rangle=\frac{1}{\sqrt{2}}\,(|g\rangle\pm |e\rangle)$, 
rather than in the $\{|g\rangle,|e\rangle\}$ basis.
We first of all rewrite $|\Psi(\tau)\rangle$ as follows:
\begin{equation}
|\Psi(\tau)\rangle
= \frac{1}{\sqrt{2}}\,(|\phi_+(\tau)\rangle|+\rangle 
+|\phi_-(\tau)\rangle|-\rangle),
\end{equation}
where we have defined
\begin{equation}
|\phi_\pm (\tau)\rangle = c_g(\tau) |\phi_g(\tau)\rangle\pm
 c_e(\tau) |\phi_e(\tau)\rangle.
\end{equation}
After the measurement of the qubit in the $\{|+\rangle,|-\rangle\}$
basis, we obtain with equal probability $p_+=p_-=\frac{1}{2}$
either the field state 
$\rho_+=|\phi_+\rangle\langle\phi_+|$ or
$\rho_-=|\phi_-\rangle\langle\phi_-|$. 
These conditional pure states have non-zero components
in the Fock basis both for even and odd number of photons and can
therefore be reconstructed by our tomographic method.
We finally obtain 
$|\phi_g\rangle=\frac{1}{2 c_g}\,(|\phi_+\rangle + |\phi_-\rangle)$ 
and 
$|\phi_e\rangle=\frac{1}{2 c_e}\,(|\phi_+\rangle - |\phi_-\rangle)$.
An example of the reconstruction of the states 
$|\phi_g\rangle$ and $|\phi_e\rangle$ is shown in 
Fig.~\ref{fig:DCE0}. We finally note that also the unconditional 
state $\rho=|c_g|^2 |\phi_g\rangle\langle \phi_g| +
|c_e|^2 |\phi_e\rangle\langle\phi_e|$ can be reconstructed,
with $p_g=|c_g|^2$ and $p_e=|c_e|^2$ probabilities of the 
outcomes $g$ and $e$ when the qubit is measured in the 
$\{|g\rangle,|e\rangle\}$ basis. 

\section{Statistical errors}
\label{sec:errors}

We now consider the realistic situation where at each discrete time 
$t_k=k(\Delta t)$ ($k=1,2,...,N_t$) 
a finite number $N_m$ of measurements is performed
and the average of the measurement outcomes is computed.
For each measurement, the system is initially prepared in the same unknown 
targed state $\rho$ and evolves (interacting with the qubit) till time
$t_k$, where the polarization of the qubit, say along the $z$-axis, is measured.
As a result, for each experimental run we obtain either $+1$ or $-1$
from the polarization measurement, with the a priori probabilities set 
by the postulates of quantum mechanics. The actually measured polarization
$z_M(t_k)$ is then the sum of two parts, the polarization 
expectation value $z(t_k)$ (recovered in the limit $N_m\to\infty$) and
the ``noise'' $z_N(t_k)$ (due to finite $N_m$), 
which can be modeled as a white noise of root mean square (RMS) amplitude 
$1/\sqrt{N_m}$. The Fourier transform of the white noise is flat and its RMS amplitude
is given by $\xi\propto 1/\sqrt{N_m}$, independently of $N_t$. On the other hand,
if we keep $\Delta t$ constant and increase both $N_t$ and 
$T=N_t(\Delta t)$, we expect that the signal (peak area in the Fourier transform) 
increases proportionally to the number of time intervals, while the noise, 
being random, increases as $\sqrt{N_t}$. Therefore the signal to
noise ration $S/\xi$ increases as $\sqrt{N_t}$. 

\begin{figure}
\begin{center}
\centerline{\includegraphics[scale=0.47]{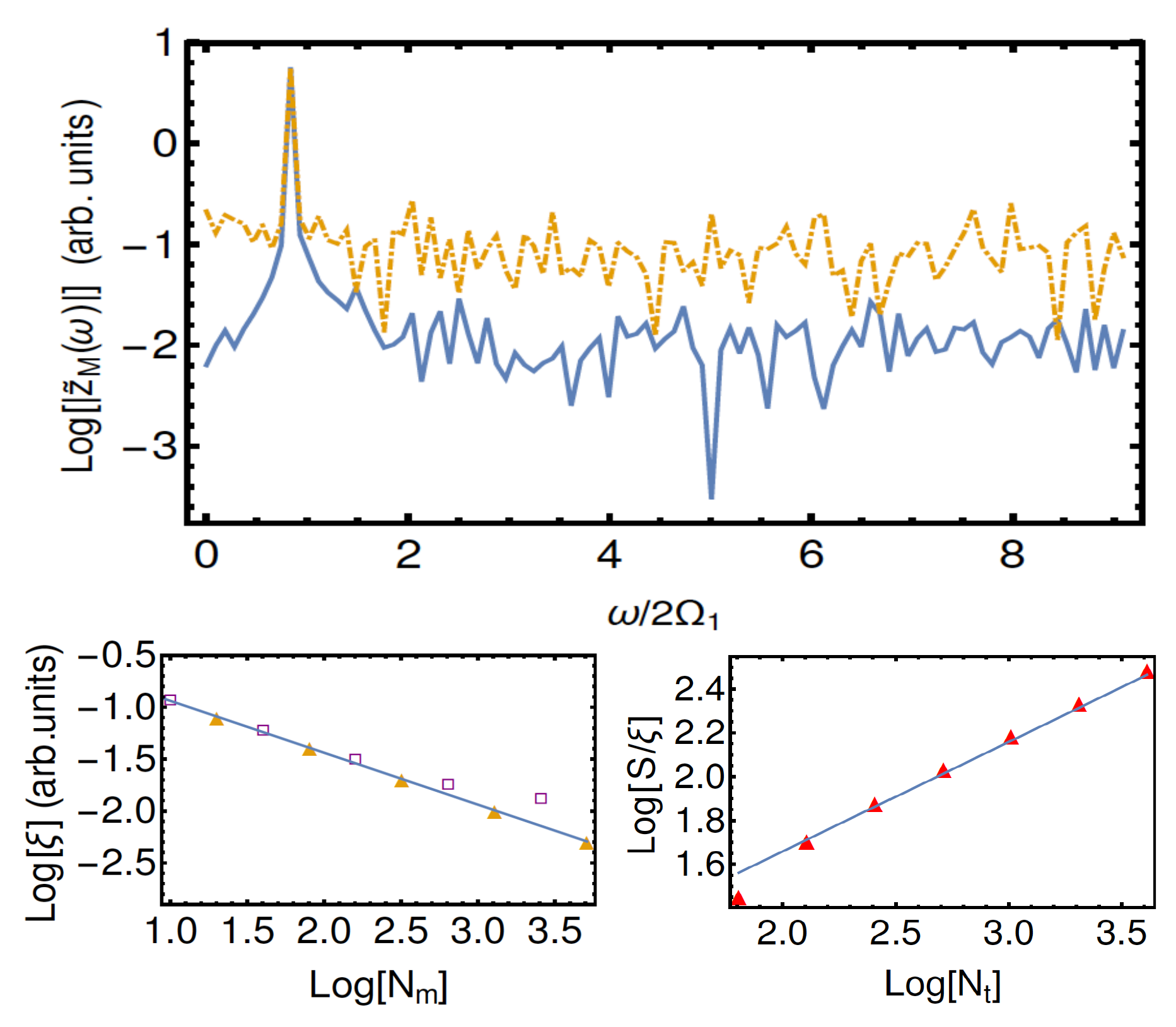}}
\caption{Top: Fourier transform of $z_M(t)$ for $T=20\pi/\Omega_1$,
$N_t=512$, and $N_m=10$ (dashed curve) and $1000$ (solid curve).
Bottom left: root mean square noise amplitude as a function of
the number of measurements $N_m$, for 
$T=20\pi/\Omega_1$, $N_t=128$ (squares) and $N_t=1024$ (triangles). 
The straight line corresponds to $\xi\propto 1/\sqrt{N_m}$
Bottom right: signal over noise ratio $S/\xi$ as function of $N_t$,
for $\Delta t=0.075/\Omega_1$, $N_m=1000$. The straight line gives the theoretical dependence
$S/\xi\propto \sqrt{N_t}$. In all panels the logarithms are base 10.}
\label{fig:statistics}
\end{center}
\end{figure}

The above expectations are confirmed by the numerical simulation of the simplest
instance of state reconstruction, namely a Fock state, here
$\rho=|1\rangle\langle 1|$. 
The Fourier transform of $z_M(t)$ shows a peak corresponding to 
$\omega=\Omega_1$ and a flat noise, see Fig.~\ref{fig:statistics} top panel 
(the symmetric peak at $\omega=-\Omega_1$ is not 
shown in the figure).
Finally, in the bottom plots of Fig.~\ref{fig:statistics} we show that the 
RMS noise amplitude $\xi\propto 1/\sqrt{N_m}$, while 
the signal over noise ratio $S/\xi\propto \sqrt{N_t}$. 
Note that in the above estimates we have assumed for the sake of simplicity
the maximum statistical error. Such error can be smaller depending on the 
signal amplitude. For instance, for a signal such that $z=+1$ (or $z=-1$)
the measurement has a constant value and statistical fluctuations vanish.

In a real experiment, decoherence effects would damp the 
oscillations in the Bloch sphere coordinates. Assuming a simple
exponential decay, $z(t)\propto e^{-\Gamma t}$, the peaks in the
Fourier transform $\tilde{z}(\omega)$ broaden but can still be resolved, 
provided the decay rate $\Gamma$ is sufficiently smaller than the separation 
$\Delta \omega$ between nearby peaks~\cite{footnote2}. 

\section{Conclusions}
\label{sec:conc}

In this paper, we have introduced a new tomographic method
for state reconstruction of a single mode of the electromagnetic 
field by means of its interaction with a probe qubit. The qubit
polarizations $x(t)$, $y(t)$, and $z(t)$ along three coordinate axes
are then measured at different times $t$ and from their Fourier 
transforms $\tilde{x}(\omega)$, $\tilde{y}(\omega)$, 
and $\tilde{z}(\omega)$ one can in general fully 
reconstruct pure states of the field and obtain partial information 
in the case of mixed states. That is, one 
can reconstruct the diagonal and the superdiagonal
of the density matrix $\rho$. 
The method could in principle be generalized, 
at the expense of a higher complexity but with a richer 
information of the state $\rho$, by using probe qudits
rather than qubits coherently interacting with the field.
Our method could also implement a simple instance of 
process state tomography. That is, assuming a field-qubit Jaynes-Cummings
interaction with unknown interaction strength $g$, one could use
the position of the peaks in the Fourier transforms 
$\tilde{x}(\omega)$, $\tilde{y}(\omega)$, and 
$\tilde{z}(\omega)$ to determine $g$. 
On a more general perspective, the method described in this paper
has some similarities with the quantum algorithm proposed by
Abrams and Lloyd~\cite{Lloyd} for finding eigenvalues and 
eigenvectors of an Hamiltonian operator. 
Both in the quantum algorithm of Ref.~\cite{Lloyd} and in the state reconstruction 
procedure described in this paper, the quantities of interest 
(the eigenvalues and the eigenvectors or the density matrix elements, 
respectively) are hidden in the time evolution of a suitable system 
and then extracted by means of the Fourier transform. 

\begin{acknowledgments}
Useful discussions with Chiara Macchiavello
and Massimiliano Sacchi are gratefully acknowledged.
\end{acknowledgments}




\begin{thebibliography}{0}
\bibitem{footnote}
An Hermitian matrix has $d^2$ independent elements, 
but the unit trace condition ${\rm Tr}(\rho)=1$ removes
one of them.
\bibitem{noteFourier}
While for the sake of simplicity in what follows we shall write
formulas for the Fourier transform of a continuous signal, 
one should more precisely consider the
discrete Fourier transform, the measurements being performed only
at discrete times $t_k=k(\Delta t)$ up to a finite interaction 
time $T=t_N=N(\Delta t)$ (see the examples of Secs.~\ref{sec:examples}
and \ref{sec:errors}).
\bibitem{footparametric}
Alternatively and with results analogous to those discussed
in this paper, the time dependence can be considered in the
detuning $\omega_a-\omega$ rather than in the coupling constant.
\bibitem{footnote2}
A more detailed analysis is required when low-frequency 
non-Markovian noise is relevant, see for instance
\cite{nmarkov1,nmarkov2,nmarkov3}.
\end{thebibliography}
\end{document}